\documentclass[review,number,sort&compress]{elsarticle}

\journal{Journal of \LaTeX\ Templates}

\begin{document}

\begin{frontmatter}

\title{Realization of a high quality factor resonator with hollow dielectric cylinders for axion searches}

\author{D. Alesini,$^1$\footnote{Corresponding author} C. Braggio,$^{2, 3}$ G. Carugno,$^{2, 3}$ N. Crescini,$^{3, 4}$ D. D' Agostino,$^5$
D. Di Gioacchino,$^1$ R. Di Vora,$^{2, 6}$ P.
Falferi,$^7$ U. Gambardella,$^5$ C. Gatti,$^1$\footnote{Corresponding author} G. Iannone,$^5$ C. Ligi,$^1$ A. Lombardi,$^4$ G. Maccarrone,$^1$ A. Ortolan,$^4$
R. Pengo,$^4$ C. Pira,$^4$ A. Rettaroli,$^{1, 8}$ G. Ruoso,$^4$, L. Taffarello,$^2$ and S. Tocci$^1$}
\address{
(QUAX Collaboration)\\
1)INFN, Laboratori Nazionali di Frascati, Frascati (Roma) Italy\\
2)INFN, Sezione di Padova, Padova, Italy\\
3)Dip. di Fisica e Astronomia, Padova, Italy\\
4)INFN, Laboratori Nazionali di Legnaro, Legnaro (PD), Italy\\
5)Dip. di Fisica E.R. Caianiello, Fisciano (SA), Italy and INFN, Sez. di Napoli, Napoli, Italy\\
6)Dip. di Fisica, Siena, Italy\\
7)Istituto di Fotonica e Nanotecnologie, CNR, INFN - TIFPA and FBK, Povo, Trento, Italy\\
8)Dip. di Matematica e Fisica Universit\`a di Roma Tre, Roma, Italy\\
}

\begin{abstract}
  The realization and characterization of a high quality factor resonator composed of two hollow-dielectric cylinders with its pseudo-TM$_{030}$ mode resonating at 10.9 GHz frequency is discussed. The quality factor was measured at the temperatures 300~K and 4~K obtaining  $\mbox{Q}_{300\mbox{K}}=(150,000\pm 2,000)$ and $\mbox{Q}_{4\mbox{K}}=(720,000\pm 10,000)$respectively, the latter corresponding to a gain of one order of magnitude with respect to a traditional copper cylindrical-cavity with the corresponding TM$_{010}$ mode resonating at the same frequency. The implications to dark-matter axion-searches with cavity experiments are discussed showing that the gain in quality factor is not spoiled by a reduced geometrical coupling $C_{030}$ of the cavity mode to the axion field. This reduction effect is estimated to be at most 20\%. Numerical simulations show that frequency tuning of several hundreds MHz is feasible.
\end{abstract}


\end{frontmatter}


\section{Introduction}

\begin{figure}[htbp]
  \begin{center}
    \includegraphics[totalheight=5cm]{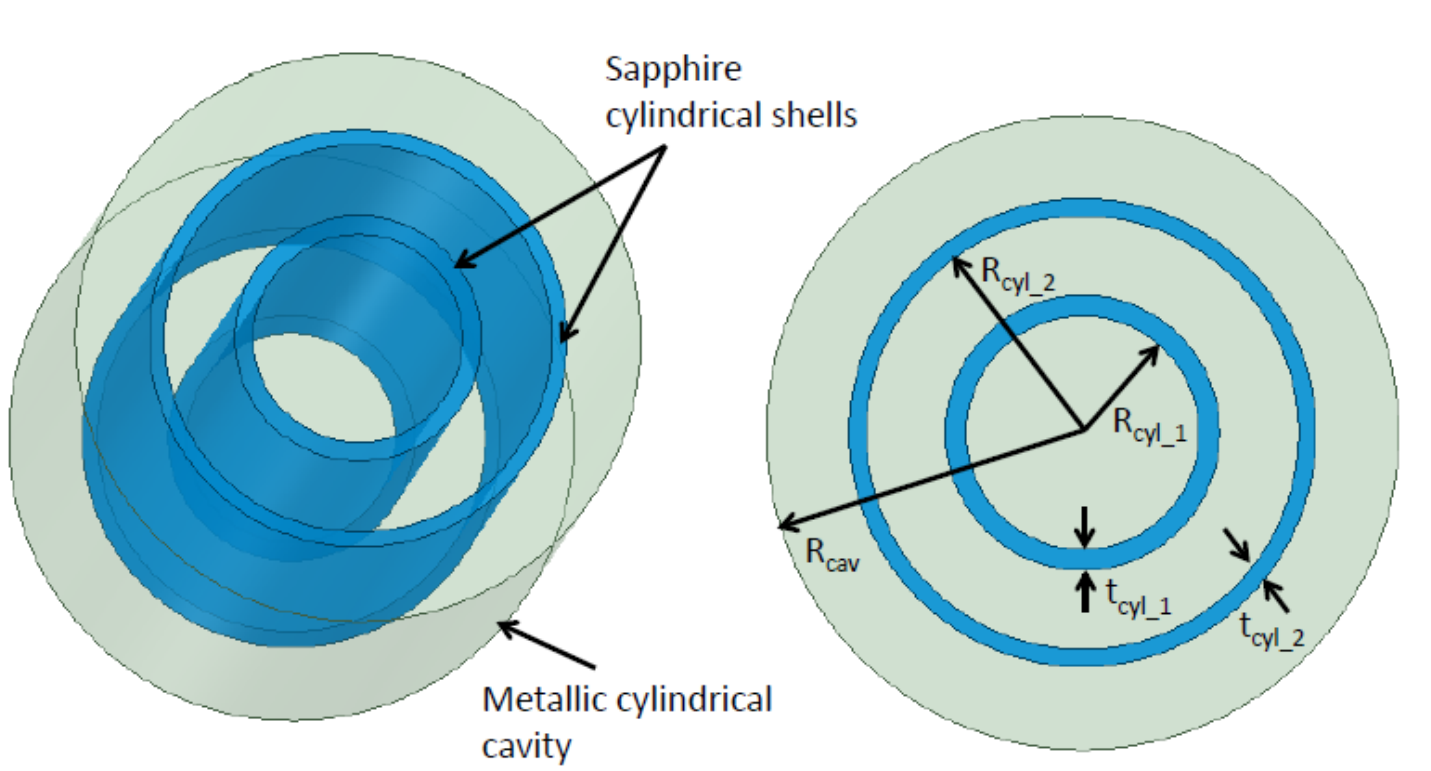}
    \caption{Sketch of the cavity with dielectric cylindrical shells.}
    \label{fig:cavitysketch}
  \end{center}
\end{figure}

The operation of the three-dimensional micro-cavities in the static magnetic field with high quality-factor plays an important role in cavity quantum-electrodinamycs experiments~\cite{Stammeier, Reshitnyk} and in the search of dark-matter (DM) axions~\cite{PecceiQuinn, PecceiQuinn2, Weinberg, Wilczek} with haloscopes~\cite{Sikivie}. While losses at high frequency limit the use of copper cavities in these experiments, an intense external magnetic-field limits the high performances achievable with superconductive cavities by causing losses due to the presence of fluxons~\cite{Abrikosov,QUAXIEEE} or a transition to the normal, resistive, state. Moreover, in quantum-electrodynamics experiments, partial or complete screening by the superconductive walls prevent controlling superconducting qubits with external magnetic fields.
Although encouraging results were obtained with NbTi and YBCO cavities~\cite{QUAXag,QUAXIEEE,CAPPYBCO},
 the possibility of using dielectric structures in axion experiments aroused considerable interest~\cite{McAllister,Caldwell,Rybka,Kim,QUAXph}.
In this paper we describe the design, fabrication and test of a pseudo-cylindrical cavity with dielectric shells made of sapphire enclosed in a copper cavity as sketched in Fig.~\ref{fig:cavitysketch} with an extremely high quality factor (potentially larger than 10$^6$ in the {\it X}-band frequency range at cryogenic temperature as shown in simulations) and with a relatively simple and tunable geometry. These types of geometries have been already implemented by other authors for different applications~\cite{Floch,Floch2,Tobar,Tobar2,Krupka2}. In particular, the realization of a resonant cavity with hollow dielectric-cylinders for axion haloscopes was first proposed in~\cite{McAllister} and a proof-of-concept demonstration was done in~\cite{Kim}. As pointed out by  these authors, if properly designed, the two cylindrical sapphire shells act as shielding that strongly reduces the magnetic field amplitude on the outer wall of the cavity and, therefore, the power losses. More precisely, the cavity working mode in this type of structures is a pseudo-TM$_{0n0}$
mode where the dielectric shells, acting as electromagnetic screens, reduce the amplitude of the secondary lobes of the field with respect to the main one. On the other hand, losses due to sapphire can be negligible thanks to its low loss-tangent, going from about 10$^{-5}$, at room temperature, down to a fraction of 10$^{-7}$ at cryogenic temperatures for purest samples~\cite{Krupka,Konaka}.

In the following we focus on the application of our dielectric cavity to DM-axion searches.
Many cavity-haloscope experiments are taking data or have been proposed in recent years: ADMX~\cite{ADMX},
HAYSTAC~\cite{HAYSTAC}, ORGAN~\cite{ORGAN}, CAPP~\cite{CAPP}, KLASH~\cite{KLASH}, RADES~\cite{RADES} and QUAX~\cite{QUAXag}.
When the resonant frequency of the cavity is tuned to the corresponding axion mass, $\nu_c=m_ac^2/h$,
the expected power generated by DM axions is given by~\cite{HAYSTAC}:
\begin{equation}
\label{eq:power}
P_{\mbox{sig}}=\left( g_{\gamma}^2\frac{\alpha^2}{\pi^2}\frac{\hbar^3 c^3\rho_a}{\Lambda^4} \right) \times
\left( \frac{\beta}{1+\beta} \omega_c \frac{1}{\mu_0} B_0^2 V C_{mnl} Q_L \right)
\end{equation}
where $\rho_a\sim0.4$~GeV/cm$^3$ is the local DM density~\cite{pdg}, $\alpha$ is the fine-structure constant, $\mu_0$ the vacuum permeability, $\Lambda=78$~MeV
is a scale parameter related to hadronic physics,
$g_{\gamma}$ the photon-axion coupling constant with central value equal to $-0.97(0.36)$ in the KSVZ (DFSZ) model~\cite{KSVZ,KSVZ2,DFSZ,DFSZ2}.
It is related to the coupling appearing in the
Lagrangian $g_{a\gamma\gamma}=(g_{\gamma}\alpha/\pi\Lambda^2)m_a$. The second parentheses contain the magnetic
field strength $B_0$, the cavity volume $V$,
its angular frequency $\omega_c=2\pi\nu_c$, the coupling between cavity and receiver $\beta$ and the loaded
quality factor $Q_L=Q_0/(1+\beta)$, where $Q_0$ is
the unloaded quality factor. $C_{mnl}$ is a geometrical factor depending on the cavity mode:
\begin{equation}
  \label{eq:cfactor}
  C_{m,n,l}=\frac{ \left| \int dV \vec{E} \cdot \vec{B_0} \right|^{2}}{ V B_0^2 \int dV \epsilon_r \left| E \right|^{2}}.
\end{equation}
where $\vec{E}$ is the electric-field mode excited by the axion field.

\section{Cavity Design}

\begin{figure}[htbp]
  \begin{center}
    \includegraphics[totalheight=5cm]{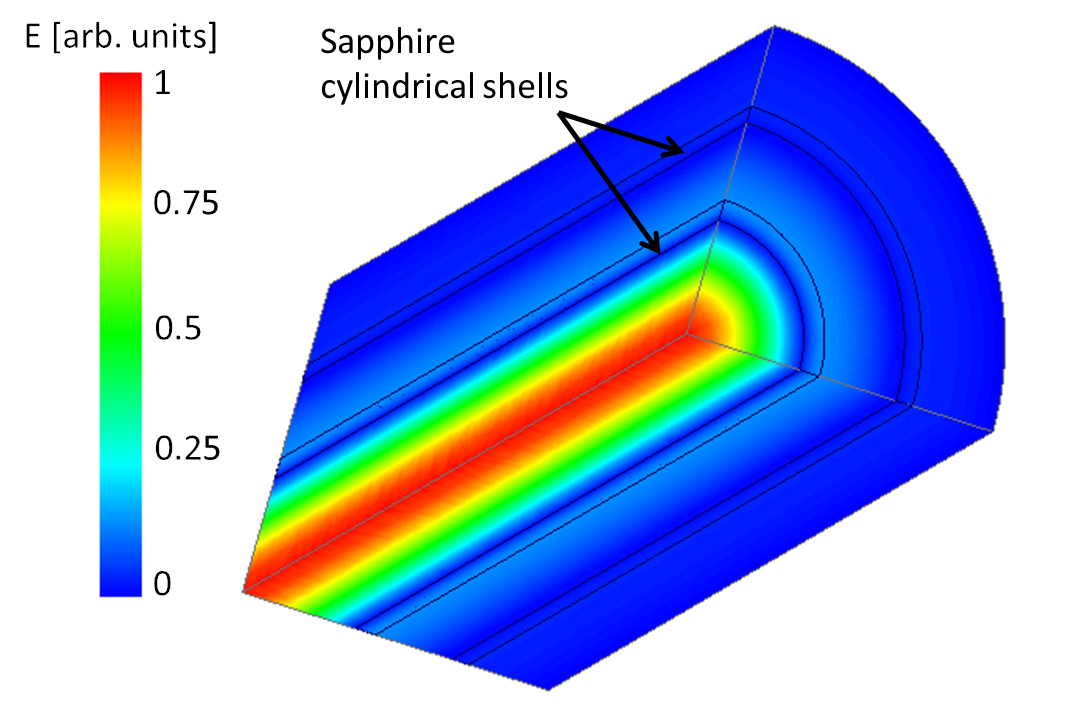}
    \caption{(colour online) Electric field amplitude of the pseudo TM$_{030}$ mode. }
    \label{fig:efield}
  \end{center}
\end{figure}

In the proposed configuration, with two shells, the selected mode is the pseudo-TM$_{030}$, as given in Fig.~\ref{fig:efield}, where we show the electric field amplitude in one quarter of the cavity. The cavity resonant-frequency {\it f}$_{res}$ was tuned to 10.9~GHz and the dielectric shells geometrical parameters
(R$_{\rm cyl\_1}$, R$_{\rm cyl\_2}$, t$_{\rm cyl\_1}$, t$_{\rm cyl\_2}$
 as defined in Fig.~\ref{fig:cavitysketch}) were optimized to minimize the losses in the outer walls. The choice of this frequency was mainly given by the possibility of incorporating the resonator inside the detection chain developed within the QUAX haloscope~\cite{QUAX}.

\begin{figure}[htbp]
  \begin{center}
    \includegraphics[totalheight=7cm]{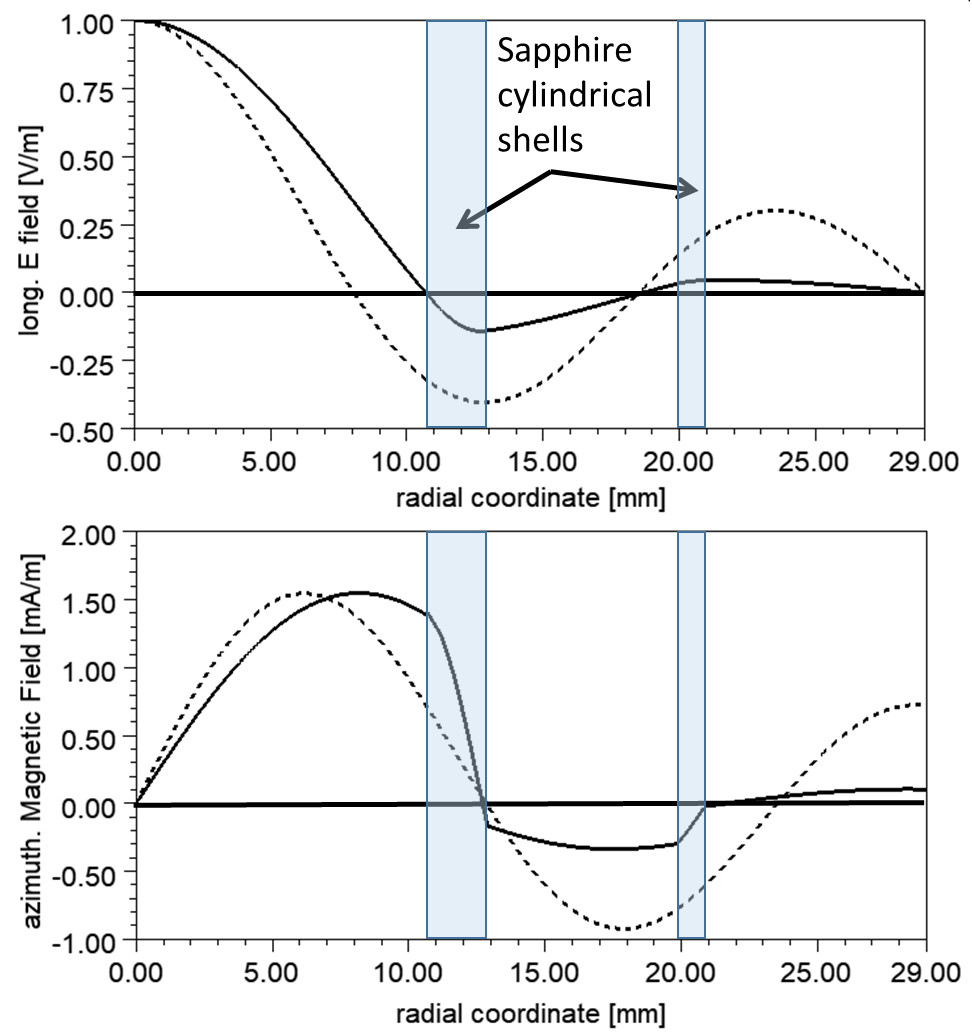}
    \caption{Longitudinal electric field and azimuthal magnetic field as a function of the transverse coordinate for the cavity with sapphire shells (continous lines) and for an ideal cylindrical cavity operating in the TM$_{030}$ mode (dotted lines).}
    \label{fig:evsx}
  \end{center}
\end{figure}
\begin{figure}[htbp]
  \begin{center}
    \includegraphics[totalheight=5cm]{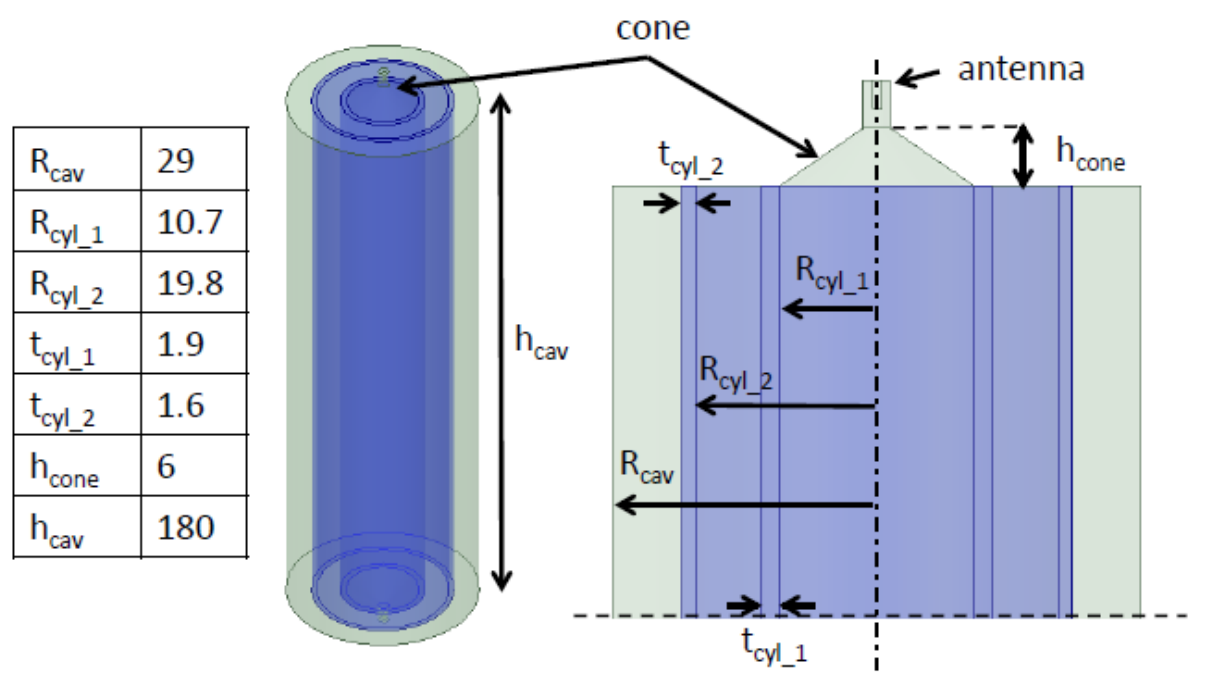}
    \caption{Final design of the fabricated cavity with its main dimensions in millimeters. Both end-plates have a conical shape to further reduce losses. See Fig.~\ref{fig:cavitytech} for details on the copper-cavity design and Fig.~\ref{fig:cavityfabrication} for the final realization.}
    \label{fig:cavitylayout}
  \end{center}
\end{figure}

The longitudinal electric field and the azimuthal magnetic field are shown in Fig.~\ref{fig:evsx} as a function of the transverse coordinate, compared with the results expected for an empty ideal cylindrical cavity operating in the TM$_{030}$ mode. The presence of the two sapphire shells reduces the amplitude of the outer field lobes and simultaneously concentrates the mode in the internal cylinder. This results in a larger form factor $C_{030}$ of the mode~\cite{Asztalos} and in reduced losses on the cavity outer-walls and therefore in a higher quality factor. The ratio of the two H fields on the outer wall of the cavity is about 9 and this gives a decrease of the losses of around 2 orders of magnitude. The final design of the cavity, optimized to reduce losses, with its main dimensions is given in Fig.~\ref{fig:cavitylayout}. The dimensions of the sapphire tubes reported here, whose measurement is described in section~\ref{sec:fabrication}, deviates slightly from the optimal values (see table~\ref{tab:sistematiche}).

We performed the electromagnetic design using the electromagnetic code ANSYS Electronics~\cite{Ansys}. Since the sapphire dielectric constant varies in a wide range depending on the crystal orientations, quality (dislocation density), impurities and temperature, we considered in the simulations a sapphire dielectric constant equal to 11.2 and a loss tangent at cryogenic temperature of $2\times10^{-6}$~\cite{Krupka,Konaka}. The copper surface resistance was set to 5.5~m$\Omega$ as expected in the anomalous regime at this frequency~\cite{Reuter}.
The two end-plates, shown in the figure~\ref{fig:cavitylayout}, were designed in order to reduce the power losses on the plates themselves. In particular, as already done in~\cite{QUAXag,QUAXIEEE}, two conical shapes were used. The length of the cones was chosen to have enough attenuation of the electromagnetic field on the cones themselves, thus reducing the losses on the copper endplates. To excite and detect the resonant modes, two coaxial antennas were inserted at the end of the cones.
\begin{figure}[htbp]
  \begin{center}
    \includegraphics[totalheight=5cm]{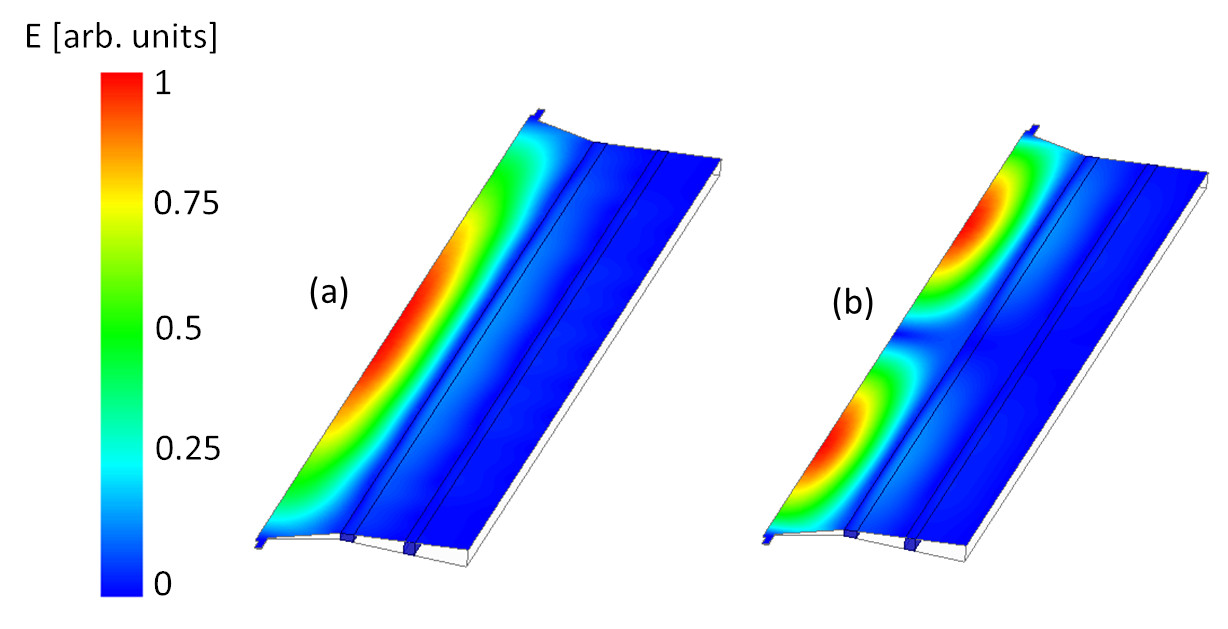}
    \caption{ (colour online) Magnitude of the electric field of the first two TM resonant modes: TM$_{030}$-like (a) and TM$_{031}$-like (b). }
    \label{fig:emag}
  \end{center}
\end{figure}

\begin{figure}[htbp]
  \begin{center}
    \includegraphics[totalheight=5cm]{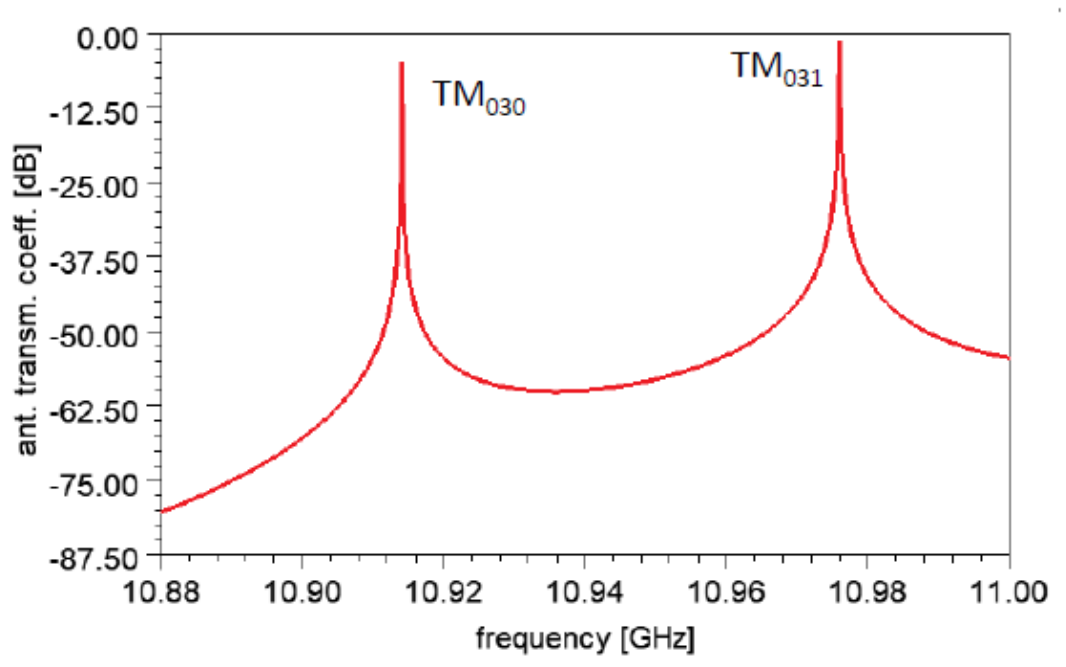}
    \caption{Simulated transmission coefficient between the two coupled antennas. The peaks correspond to the resonant modes TM$_{030}$ and TM$_{031}$.}
    \label{fig:s12}
  \end{center}
\end{figure}

The simulated magnitude of the electric field of the two TM resonant modes (TM$_{030}$-like and TM$_{031}$-like) is shown in Fig.~\ref{fig:emag}. Because of the cavity symmetry, we simulated a small sector of the cavity with perfect magnetic boundary conditions. The first mode is expected to have a Q-factor of about $1.9\times 10^6$ at cryogenic temperature, essentially limited by the losses on the cavity walls and endplates. Fig.~\ref{fig:s12} shows for completeness the simulated transmission coefficient between the two coupled antennas whose peaks correspond to the two mentioned resonant modes.
In table~\ref{tab:sistematiche} we report the variation of quality factor and resonant frequency obtained by changing some of the simulation parameters. Significant variations of the quality factor are observed as a function of the loss tangent and of the thickness of the outer sapphire tube. As discussed in section~\ref{sec:fabrication}, this second effect places precise requirements for the manufacture of sapphire tubes.

\begin{table}[h!]
  \begin{center}
    \caption{Dependence of quality factor and frequency on simulation parameters. Reference values are $Q=1.91\times10^6$ and {\it f}$_{res}$=10.915 GHz.}
    \label{tab:sistematiche}
  \vspace*{0.5cm}
    \begin{tabular}{c|c|c|c|c}
      \hline\hline
      parameter & reference value & simulated value &	 {\it f}$_{res}$ [GHz]	& Q ($10^6$) \\\hline
      Loss tangent    & $2\times10^{-6}$& $10^{-5}$	& 10.915	&  0.54 \\
      Loss tangent    & $2\times10^{-6}$& $10^{-7}$	& 10.916	&  4.98 \\
      R$_{\rm cyl\_1}$&	10.7~mm  & 10.9~mm    & 10.77   &  1.92	\\
      R$_{\rm cyl\_2}$&	19.8~mm  &	20.0~mm   & 10.902  &  1.94 \\
      t$_{\rm cyl\_1}$&	1.9~mm   &	2.1~mm    & 10.77   &  1.95 \\
      t$_{\rm cyl\_2}$ &	1.6~mm   &	1.8~mm    & 10.90   &  0.86 \\
      $\epsilon_r$    & 11.2     &	9.39      & 11.15   &  1.30 \\
      $\epsilon_r$    & 11.2     &	11.5      & 10.89   &  1.95 \\
    \end{tabular}
  \end{center}
\end{table}

We calculated, for several configurations, the ratio $R_{CV}$ of the product $C_{030}\times V$ between the form factor and the volume of the dielectric cavity and the product $C_{010}\times V$ for an ideal cylindrical-cavity of the same length with the TM$_{010}$ mode resonating at the same frequency. According to equation~\ref{eq:power} the gain in signal power is proportional to the ratio $R_{CV}$ and to the ratio of quality factors. For a cavity without cones, closed with flat endplates and sapphires geometry as reported in Fig.~\ref{fig:cavitylayout}, we obtained $R_{CV}=72\%$. The decrease is due to the field on the second lobe in figure~\ref{fig:evsx} that has an opposite sign with respect to the main one. For a cavity with the conic endplates, this value decreases depending on the length of the cones themselves with respect to the volume of the body. With our design parameters it reaches the value 66\%.
We veryfied that with optimized dimensions of the sapphire cylinders (for instance  t$_{\rm cyl\_1}$=2.2 mm and t$_{\rm cyl\_2}$=1 mm) the mentioned ratio of 72\% increases up to almost 100\%, and for  a cavity with conic endplates it increases up to 90\%. Cavities with such dimensions will be implemented in future realizations.

\begin{figure}[htbp]
  \begin{center}
    \includegraphics[totalheight=5cm]{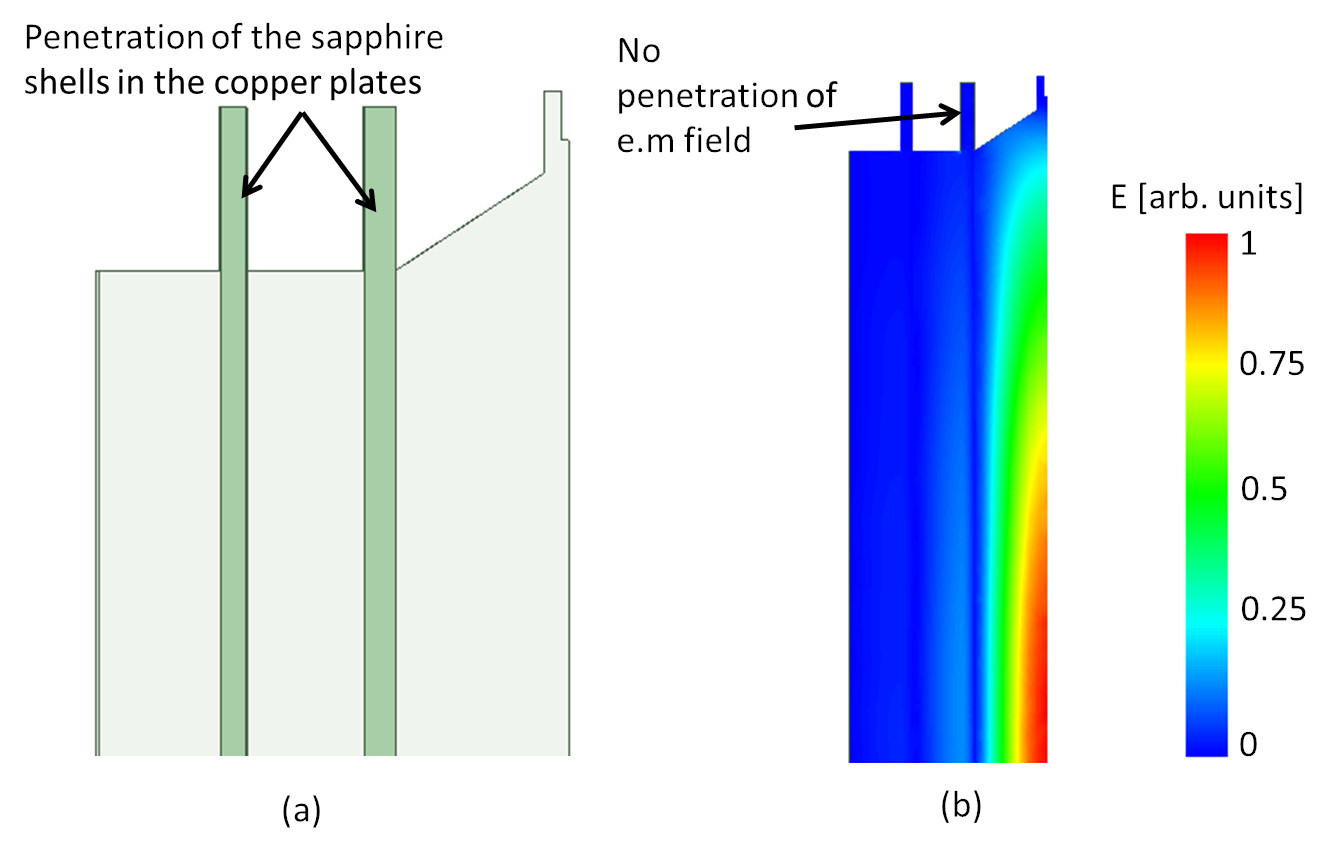}
    \caption{(colour online) Sketch of the sapphire shells penetrating into the copper material. }
    \label{fig:penetration}
  \end{center}
\end{figure}

As discussed in the next paragraph the cavity was fabricated fixing the two shells on the two endplates. More in detail, the sapphire shells penetrate into the copper for about 10~mm and are fixed to it. The sketch of the geometry is given in the left panel of Fig.~\ref{fig:penetration} while the technical drawing of the copper cavity is shown in Fig.~\ref{fig:cavitytech}. To calculate the effect of these penetrations, we simulated the whole structure including the insertion of the sapphire inside the grooves of the copper endplates. The result is given in the right panel of Fig.~\ref{fig:penetration} where we show the magnitude of the E field. The figure clearly shows that the electromagnetic field hardly penetrates into the copper-plates hollows and, as a consequence, the frequency changes by less than 0.1 permil and quality factor by less than one percent.

\begin{figure}[htbp]
  \begin{center}
    \includegraphics[totalheight=6cm]{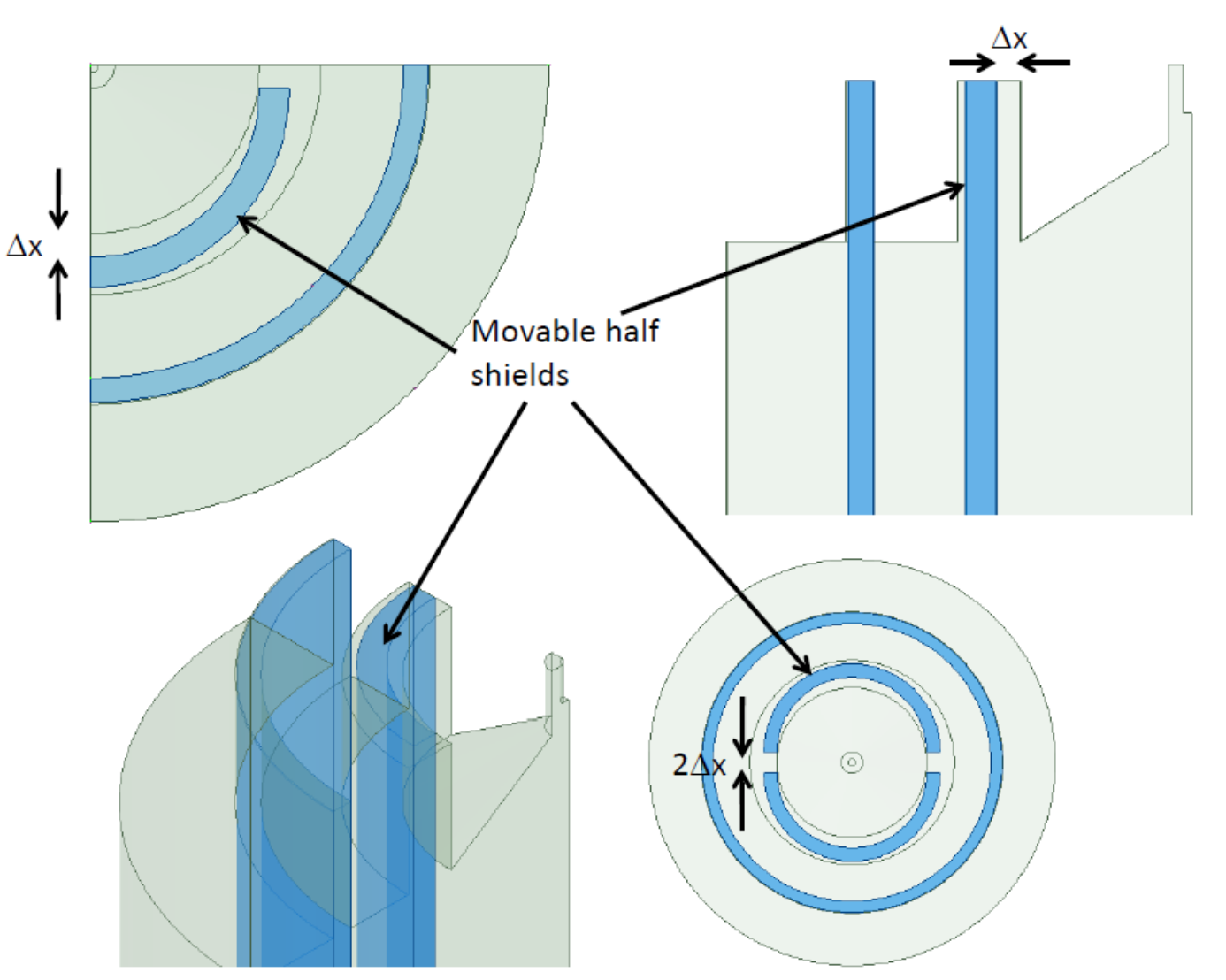}
    \caption{Sketch of the geometry showing the possible implementation of a tuning system to change the resonance frequency of the cavity without affecting its performances.}
    \label{fig:tuning}
  \end{center}
\end{figure}

\begin{figure}[htbp]
  \begin{center}
    \includegraphics[totalheight=5cm]{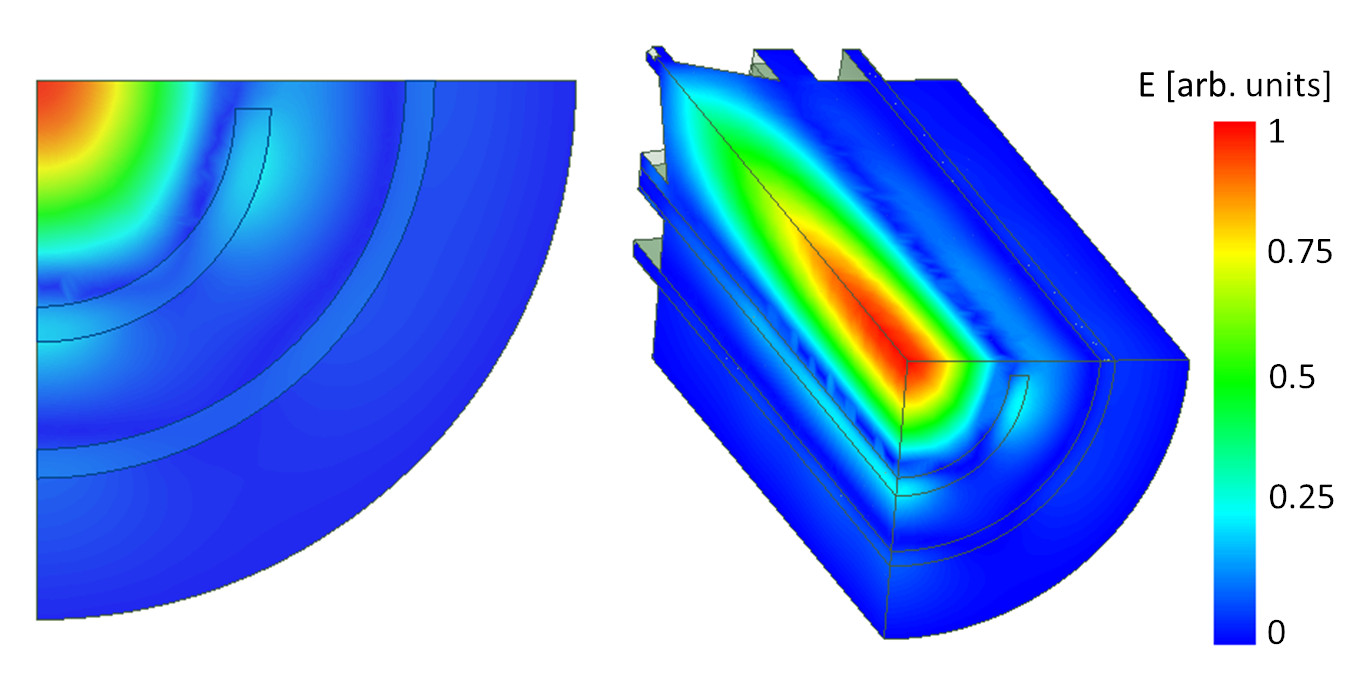}
    \caption{(colour online) Magnitude of the electric field when the distance between the two halves is 3~mm. }
    \label{fig:Etuning}
  \end{center}
\end{figure}

This type of geometry allows the implementation of a tuning system to change the resonance frequency of the cavity. The sapphire shells can be cut in two halves, as suggested in~\cite{Kim}, and the two half-cylinders can be moved by means of a mechanism embedded into the copper plates. The geometry is sketched in Fig.~\ref{fig:tuning}. The resonant frequency, the quality factor, and the factor $C\times V_{030}$ are given in table~\ref{tab:valori} for different position of the two half cylinders. The result suggests that it is possible to tune the frequency in a range of more than 500~MHz without deteriorating the performance of the cavity too much. The magnitude of the electric field when the separation between the two halves is 3~mm is given in Fig.~\ref{fig:Etuning} that clearly shows that the electromagnetic field is well contained within the two half-shells. The mechanical design of such a mechanism is still under development.

\begin{table}[h!]
  \begin{center}
    \caption{Expected frequency and quality factor for different distance $\Delta x$ between the two half cylinders. The small variation of the value of Q for $\Delta x=0$ with respect to the value in table~\ref{tab:sistematiche} is due to the different design of the cavity shown in figure~\ref{fig:tuning}.}
    \label{tab:valori}
  \vspace*{0.5cm}
    \begin{tabular}{c|c|c|c|c}
      \hline\hline
      $\Delta x$ [mm] &	 {\it f}$_{res}$ [GHz]	& Q (10$^6$) &	$C_{030}\times V$ (10$^{-6}$~m$^3$) &	$C_{030}\times V\times Q$ (m$^3$)\\\hline
      0	& 10.92	& 2.01	& 24.75 &	49.7 \\
      0.25&	10.81&	1.766&	26.23&	46.3 \\
      0.5&	10.71&	1.80&	27.62&	49.7\\
      0.75&	10.62&	1.69&	28.94&	48.9\\
      1&	10.53&	1.49&	30.00&	44.7\\
      1.25&	10.45&	1.39&	31.31&	43.5\\
      1.5&	10.38&	1.39&	32.16&	44.7\\
    \end{tabular}
  \end{center}
\end{table}

\section{Cavity fabrication and mechanical tolerance}
\label{sec:fabrication}

Two fine-grid sapphire-tubes 200~mm long were purchased from ROSTOX-N~\cite{rostox}. Their optical axes (C-axis, 0001) are oriented along the cylinder axis of symmetry within one degree, as stated by the manufacturer. We measured, with 4 points per side, the tubes diameters with a coordinate-measuring device at Laboratori Nazionali di Frascati: the smaller one has inner diameter 21.36(10)~mm and outer diameter 25.17(1)~mm; the larger one has inner diameter 39.71(4)~mm and outer diameter 42.80(1)~mm. The errors reflect the machine precision (about 10$\mu$m) or, if larger, the roundness, the maximal variation in the values measured on both sides of the tubes. We measured eccentricities, defined as the distance between the centers of inner and outer walls of a cylinder, up to about 0.2~mm. The simulation results in table~\ref{tab:sistematiche} show that although the quality factor is not sensitive to most of these small variations, an eccentricity of 0.2~mm of the larger tube, resulting in a similar variation of the tube thickness, can reduce the quality factor by a factor $\sim 2$.

\begin{figure}[htbp]
  \begin{center}
     \includegraphics[totalheight=10cm]{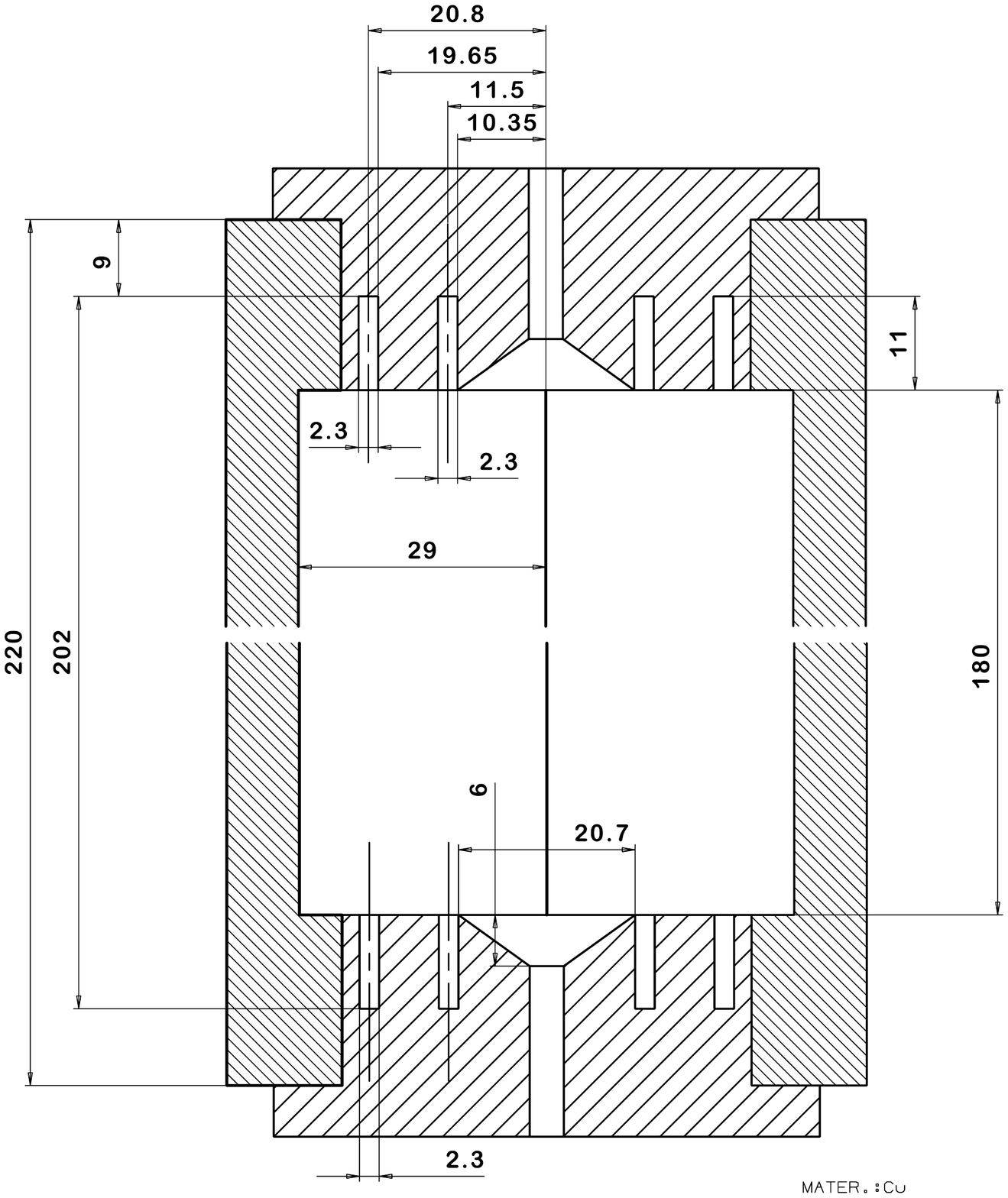}
    \caption{Technical drawing of the copper cavity housing the two sapphire cylinders. Through holes for assembly and venting holes are not shown}
    \label{fig:cavitytech}
  \end{center}
\end{figure}

The tubes were sent to Laboratori Nazionali di Legnaro (LNL) for assembly inside the copper cavity. The technical drawing of the copper cavity housing the two sapphire cylinders is shown in Fig. \ref{fig:cavitytech}. It is made of four pieces: two end caps and two lateral half sides. On the internal side of each end cap two circular grooves are carved to hold the sapphire cylinders in place. Each groove width is such so as to avoid compression of the sapphire from the copper when cooling. For the same reason, the depth of the grooves is 1~mm bigger than the designed sapphire penetration depth. The inner cylindrical volume is formed by combining the two side halves. The resulting cylinder has an inner radius of 29~mm, while from the outside the structure has a rectangular section. Three 1~mm diameter venting holes are also drilled on one end cap for every separate volume that is formed in the interior of the assembly. All the copper parts were polished by chemical etching before being mounted.

When the two lateral sides are joint together and blocked with M5 non magnetic stainless steel (AISI 304) bolts, the two end caps can still be moved indipendently and fastened. This allows for the easy positioning of the sapphire cylinders while keeping one end cap out. The cavity is normally operated keeping its axis vertical. For this reason no specific holding system  for the sapphire tubes has been designed yet, and so they just keep their positions through gravity.

\begin{figure}[htbp]
  \begin{center}
    \includegraphics[totalheight=5cm]{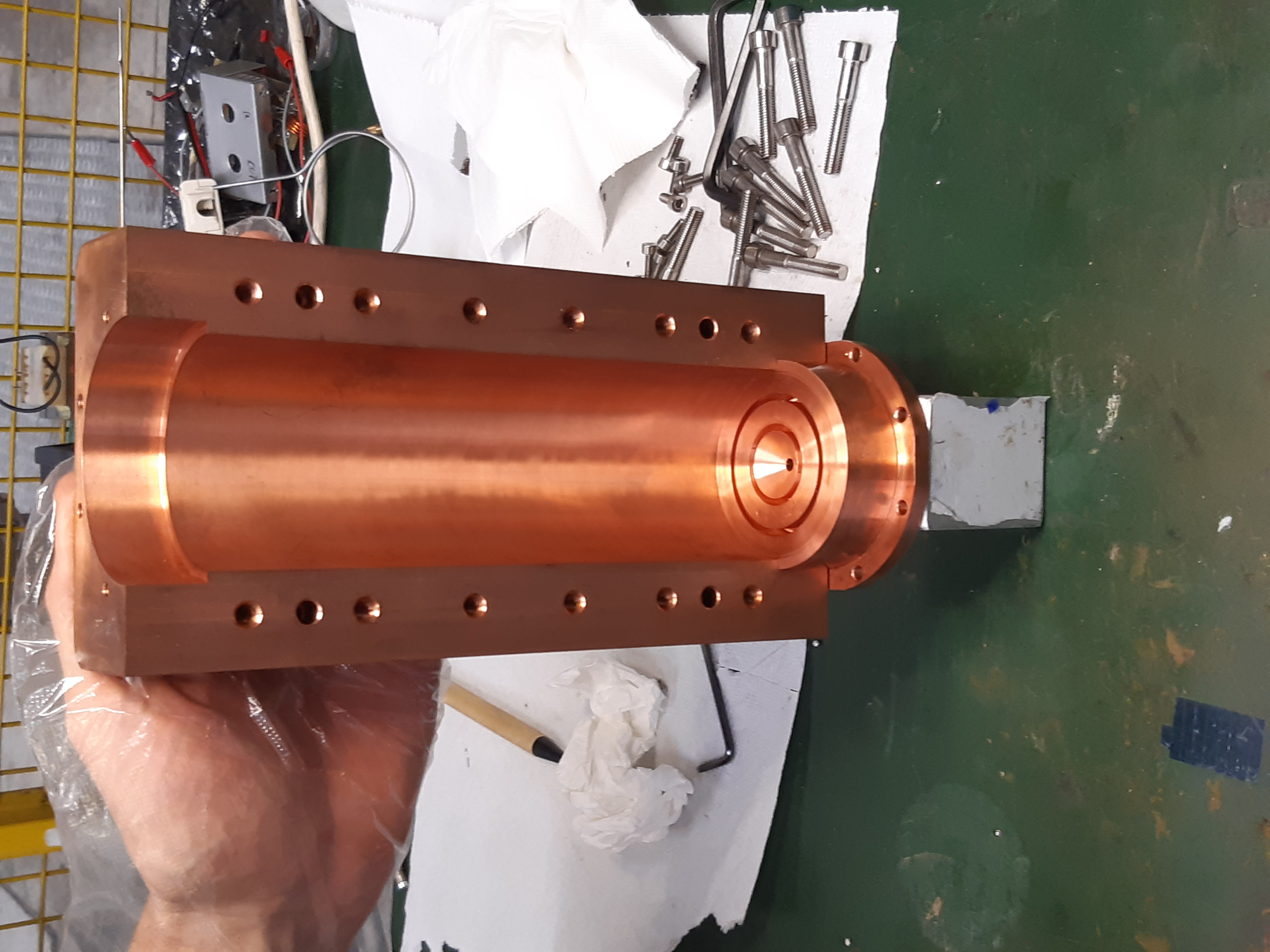}
    \includegraphics[totalheight=5cm]{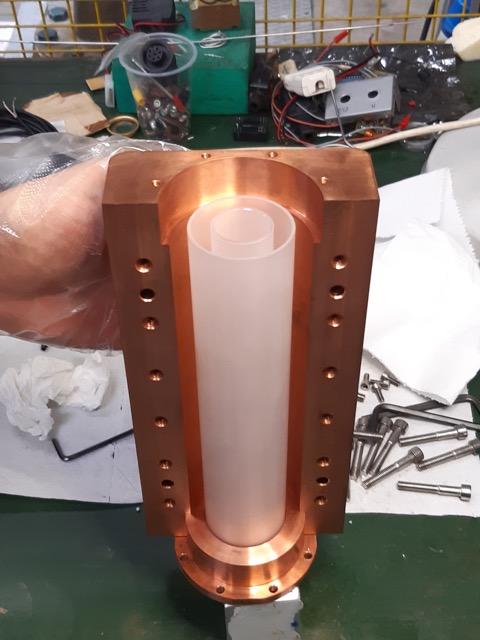}
    \includegraphics[totalheight=5cm]{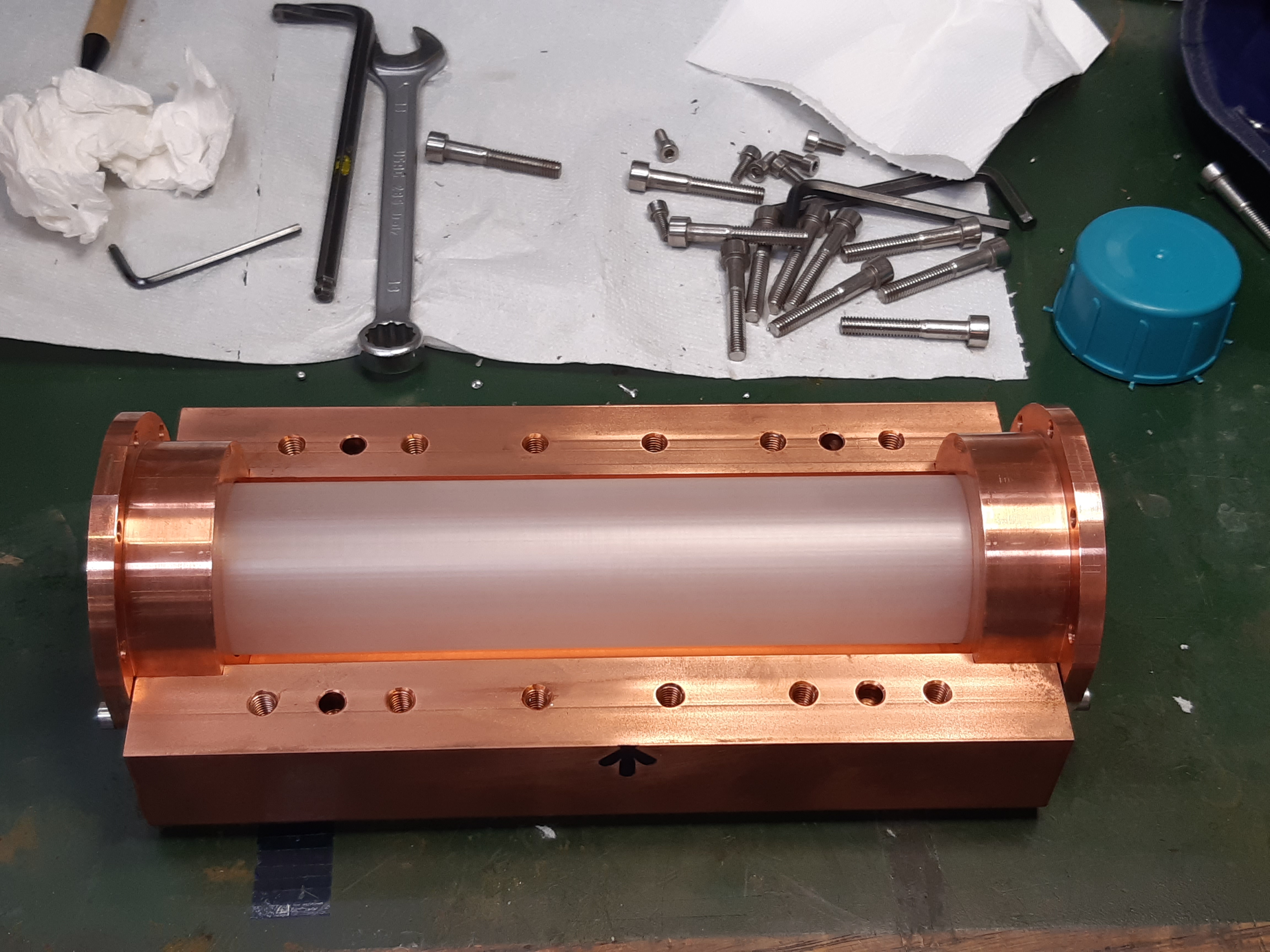}
    \caption{Partial assemblies of the dielectric cavity.}
    \label{fig:cavityfabrication}
  \end{center}
\end{figure}

Three  photographs of partial assemblies of the dielectric cavity are shown in Fig.~\ref{fig:cavityfabrication}. The final assembly of the sapphire shells is done with only one end-cap removed and the cavity in the vertical position.

\section{Experimental characterization of dielectric cavity resonant modes}

\begin{figure}[htbp]
  \begin{center}
    \includegraphics[totalheight=7cm]{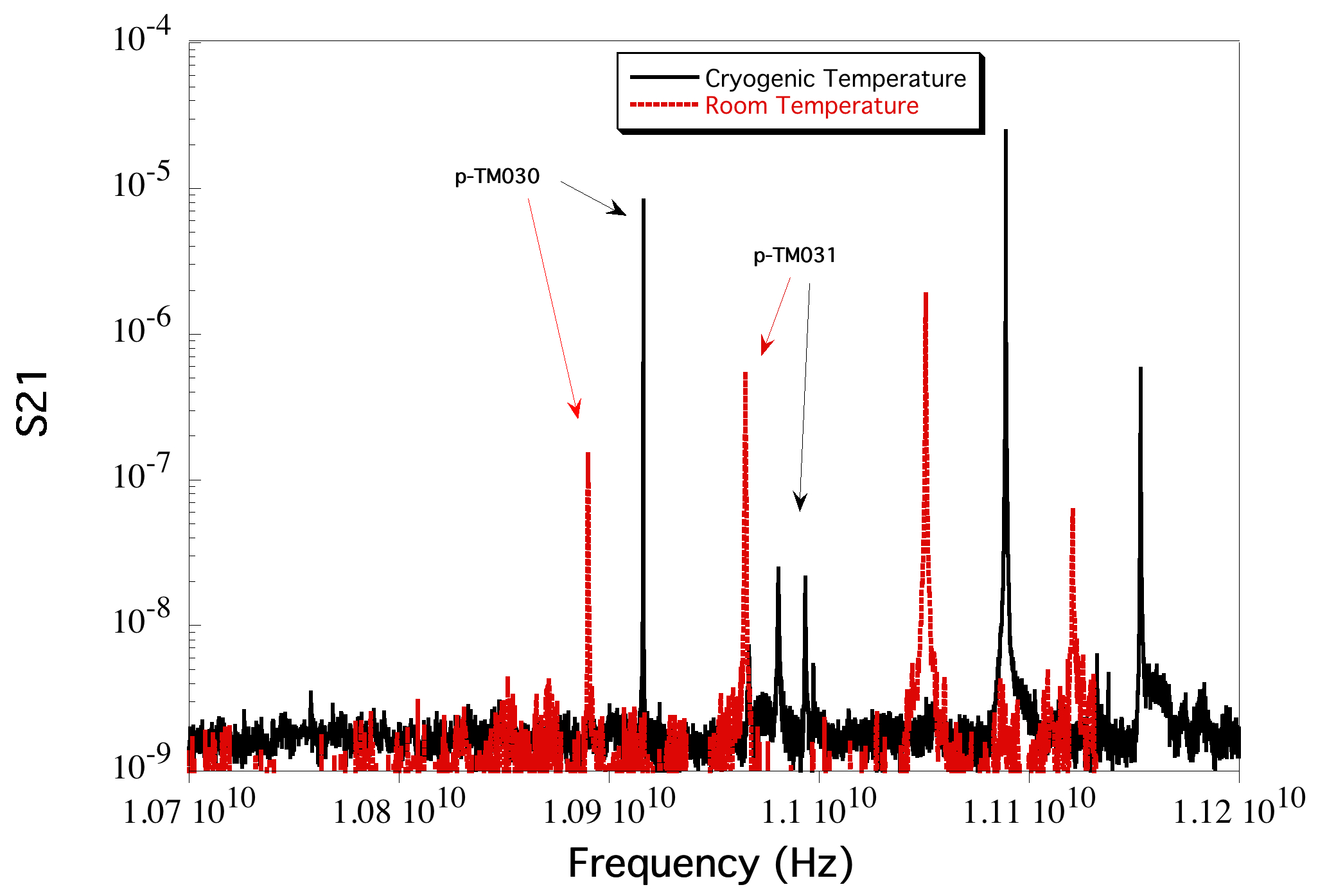}
    \caption{Measured spectrum of the resonant modes of the dielectric cavity at 5.4~K (continous line) and 300~K (dashed line). The pseudo-TM$_{030}$ is the lowest-frequency peak, among the strong ones, and the the pseudo-TM$_{031}$ is the subsequent one.}
    \label{fig:misuramodi}
  \end{center}
\end{figure}

\begin{figure}[htbp]
  \begin{center}
    \includegraphics[totalheight=6cm]{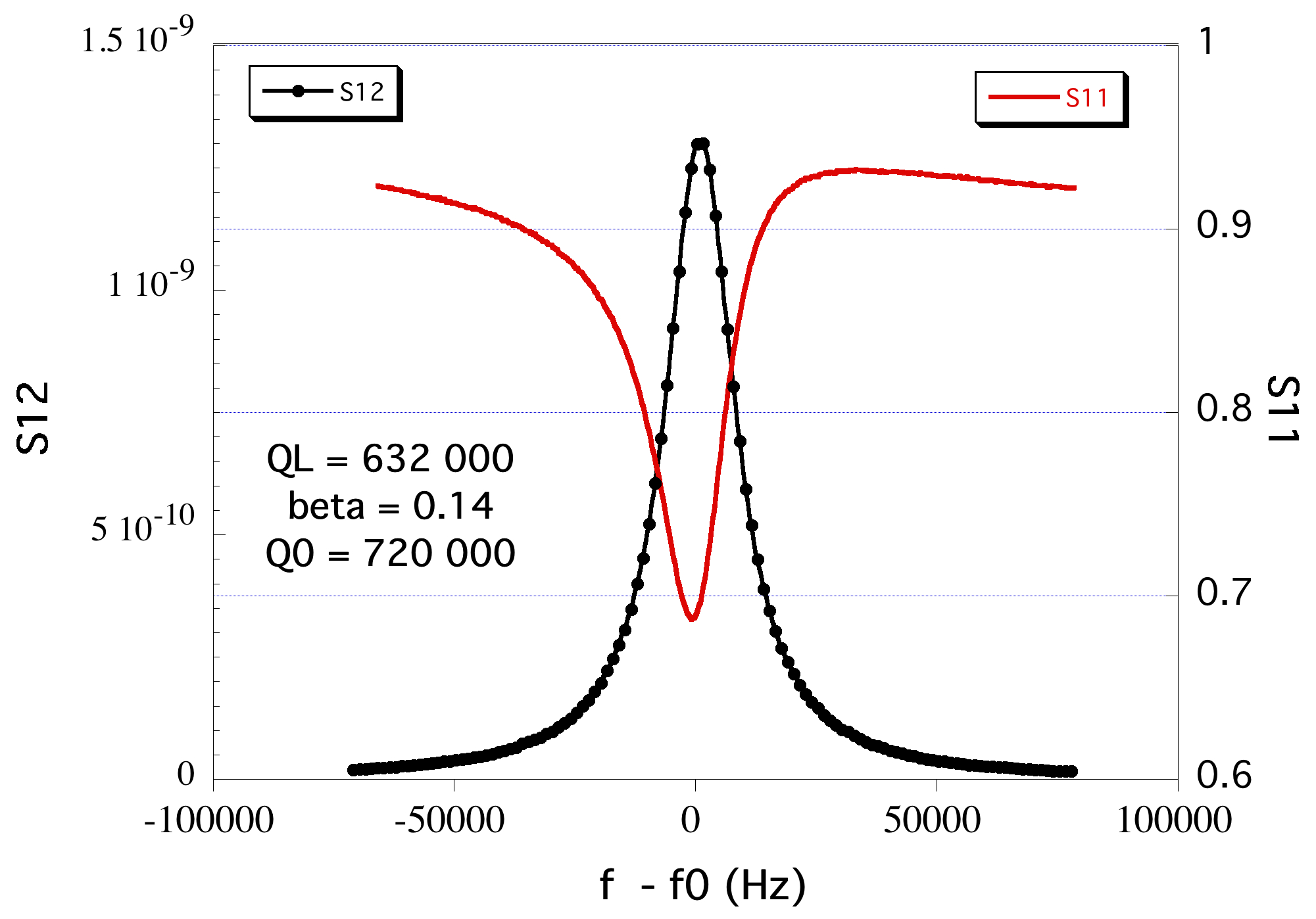}
    \caption{Trasmission and reflection parameters as a function of frequency for the pseudo-TM$_{030}$ mode at 10.916~GHz at 5.4~K.}
    \label{fig:misuraQ}
  \end{center}
\end{figure}

We characterized the resonant modes of the dielectric cavity at LNL in a LHe-cryostat at 5.4~K. We connected the cavity to two fixed antennas subcritically coupled to the pseudo-TM$_{030}$ mode and placed it inside a vacuum chamber designed to allow operation inside cryogenic dewars. The vacuum chamber is equipped with two rf feedthroughs and a thermometer measuring the temperature of the cavity. The thermometer was mounted on the external surface of the resonant cavity, thermally coupled to it. The temperature scale was calibrated within 1~K. The spectrum of the resonant modes, measured  at 300 and 5.4~K with a Vector Network Analyzer (VNA), is shown in Fig.~\ref{fig:misuramodi}. We select the pseudo-TM$_{030}$ as the lowest-frequency peak with strong coupling with the dipole antenna. At room temperature the pseudo-TM$_{030}$ mode has a frequency $\nu_{030}=10.886$~GHz with quality factor $Q_{030}=(150,000\pm 2,000)$ where the uncertainty is estimated from repeated measurements. The second peak at higher frequency corresponds to the TM$_{031}$ mode. The observed mode-frequencies are in agreement with the values obtained in the simulated spectrum shown in Fig.~\ref{fig:s12}, corresponding to 10.915~GHz and 10.975~GHz for the modes  TM$_{030}$ and TM$_{031}$, respectively.
A low exchange-gas pressure, about 10$^{-5}$~mbar, guaranteed slow cooling, preventing damages due to differential contractions of copper and sapphire. Because of thermal contractions, changes in the positions of the sapphire tubes in their housings and variation of the sapphire dielectric constant, we observed drifts and crossings of modes which were however followed by continous measurement of the transmission spectrum.
At 40~K the mode frequency reached a plateau at $\nu_{030}=10.916$~GHz with quality factor $Q_{030}=320,000$. We then added few mbar of He gas to speed up the cooling that soon stopped at 5.4~K. Transmission and reflection parameters taken at this temperature are shown in Fig.~\ref{fig:misuraQ} as measured from the port with higher coupling to the cavity, while on the other port the reflected signal was barely visible. The measured loaded quality factor is $Q_{L}=632,000$. The unloaded quality factor is calculated as $Q_{030}=(1+k)\times Q_L$ where
$k\sim (1-S_{11}(\nu_{030}))/(1+S_{11}(\nu_{030}))$ is the coupling to the antenna. We obtain $Q_{030}=(720,000\pm 10,000)$ a very large quality factor if compared with copper cavities at these frequencies and temperatures with typical quality factor of less than 100,000. The error reflects the stability of the measurement in time and the uncertainty on the determination of the coupling $k$. We also observed a 10\% variation by moving the whole cryostat in order to change the alignment of the sapphire tubes. However, missing a mechanical-movement system in our setup, we could not investigate this effect further and we postpone it until the movement system will be available. The measured quality factor and frequency are summarized in table~\ref{tab:misure}. Among the causes that may have reduced the quality factor estimated in our simulations are the 0.2~mm eccentricity of the larger tube, as discussed in section~\ref{sec:fabrication}, or to higher losses in the sapphire due to a loss tangent equal to $7.2\times10^{-6}$. This value is larger than that observed in~\cite{Krupka} but compatible with the measurement in~\cite{Konaka}.

\begin{table}[h!]
  \begin{center}
    \caption{Measured frequency and quality factor.}
    \label{tab:misure}
  \vspace*{0.5cm}
    \begin{tabular}{c|c|c}
      \hline\hline
      T & $\nu$ & Q \\\hline
      300~K &  10.886~GHz & $(150,000\pm 2,000)$ \\
      5.4~K & 10.916~GHz & $(720,000\pm 10,000)$ \\
    \end{tabular}
  \end{center}
\end{table}

\section{Conclusions}
We realized a dielectric resonance cavity composed of two concentric sapphire hollow-tubes housed in a copper cavity. Placing the sapphire tubes close to the nodes of the TM$_{030}$ mode reduces by an order of magnitude the azimuthal component of the magnetic field on the cylindrical copper wall, reducing the losses and increasing the quality factor of the pseudo-TM$_{030}$ mode resonating at 10.9 GHz up to 720,000 at a temperature of 5.4~K. Electromagnetic simulations show that the frequency mode is in principle tunable in a 500~MHz range. This result improves the one previously obtained by our group with a NbTi cavity~\cite{QUAXag} and more importantly, we expected it to be unaffected by an applied magnetic field. This quality factor is close to $10^{6}$, the limit  imposed by the signal linewidth expected from DM axions, and is expected to further improve, up to $2\times10^{6}$, by properly tuning the thickness of the sapphire tubes or by using sapphire tubes with lower loss tangent.

\section{Aknowledgments}

We are grateful to E. Berto, A. Benato and M. Rebeschini,  who did the mechanical work,  F. Calaon and M. Tessaro who helped with the electronics and cryogenics, and  to  F.  Stivanello  for  the  chemical  treatments.   We thank M. Zago who realized the technical drawings of the system. We deeply acknowledge the Cryogenic Service of the Laboratori Nazionali di Legnaro, for providing us the  liquid helium. We also acknowledge M. Matteo and E. Dan\`e of the Metrological Service of Laboratori Nazionali di Frascati for the precise measurement and analysis of the sapphire cylinders.

\section*{References}

\bibliography{mybibfile}

\end{document}